\pdfoutput=1
\documentclass[a4paper]{jpconf}
\usepackage{graphicx}
\usepackage[T1]{fontenc}
\usepackage[latin1]{inputenc}
\usepackage[thinspace,thinqspace]{SIunits}					

\newcommand{\YRS}{YbRh$_2$Si$_2$}
\newcommand{\YIyS}{Yb(Rh$_{1-y}$Ir$_y$)$_2$Si$_2$}
\newcommand{\YIsS}{Yb(Rh$_{0.94}$Ir$_{0.06}$)$_2$Si$_2$}
\newcommand{\YCxS}{Yb(Rh$_{1-x}$Co$_{x}$)$_2$Si$_2$}
\newcommand{\YCdS}{Yb(Rh$_{0.97}$Co$_{0.03}$)$_2$Si$_2$}

\newcommand{\TFL}{\ensuremath{T_{\mathrm{LFL}}}}									
\newcommand{\TN}{\ensuremath{T_{\mathrm{N}}}}									
\newcommand{\HN}{\ensuremath{H_{\mathrm{N}}}}									
\newcommand{\Tstar}{\ensuremath{T^{\star}}}										
\newcommand{\Hstar}{\ensuremath{H^{\star}}}										
\newcommand{\percent}{\%}																
\newcommand{\mT}{\milli\tesla}														
\newcommand{\Neel}{N\'eel}																
\newcommand{\HcA}{\ensuremath{H_{\mathrm c}^A}}									
\newcommand{\dd}{\mathrm{d}} 															
\newcommand{\mK}{\milli\kelvin}

\begin{document}
\title{Quantum criticality in \YCdS\ probed by low-temperature resistivity}

\author{Sven Friedemann, Niels Oeschler, Cornelius Krellner, Christoph Geibel and Frank Steglich}

\address{Max Planck Institute for Chemical Physics of Solids, 
01187 Dresden, Germany}

\ead{Sven.Friedemann@cpfs.mpg.de}

%
%
\begin{abstract}
Quantum criticality in \YCdS\ is investigated by means of resistivity and magnetoresistance. The partial substitution of Co leads to a stabilization of the magnetism as expected according to the application of chemical pressure for Yb systems. However, the signature of the Kondo-breakdown remains at the same position in the temperature-magnetic field phase diagram compared to stoichiometric \YRS. As a consequence, the Kondo-breakdown is situated within the antiferromagnetic phase. These results fit well within the global phase diagram of \YRS\  under chemical pressure \cite{Friedemann2009}.
\end{abstract}

%
%
\section{Introduction}
Considerable effort has recently been devoted to the investigation of quantum critical points (QCPs) which arise when a continuous phase transition is driven to zero temperature by a non-thermal parameter \cite{Focus2008}. 
In heavy fermion systems (HFS), the Kondo effect leads to the formation of composite quasiparticles of the $f$ and conduction electron states with huge effective masses. 
At present two main theoretical scenarios describe the antiferromagnetic (AF) QCP in HFSs. 
The conventional one 
being applicable to many HFSs assumes that the quasiparticles stay intact \cite{HMM}. 
However, a series of unconventional theoretical descriptions discard this basic assumption \cite{SPS}. Rather, they focus on the breakdown of the Kondo effect causing the $f$ and conduction electrons to decouple. 
 As a consequence the Fermi surface may reconstruct, in contrast to the smooth evolution expected in the conventional case \cite{Coleman2001}. A new energy scale \Tstar\ is predicted reflecting the finite temperature crossover of the Fermi-surface volume. This picture seems to apply to \YRS, a very clean HFS ideally suited to investigate quantum criticality \cite{CT}: AF order sets in at $\TN = \unit{0.07}\kelvin$ and can easily be suppressed by a small magnetic field of $\mu_0\HN = \unit{60}\mT$ applied perpendicular to the crystallographic $c$ axis. A rapid change of the Hall coefficient was detected along a line $\Tstar(H)$ which merges with \HN, the width of the Hall crossover extrapolating to zero for $T \to 0$ \cite{Paschen2004}. This evidences an abrupt change of the FS indicating a correspondence between $\Tstar(H)$ and the Kondo-breakdown (KB) energy scale. Subsequent thermodynamic and transport investigations confirmed $\Tstar(H)$ to be a new energy scale \cite{Gegenwart2007}. One signature of \Tstar\ was observed in the magnetoresistance which exhibits a S-shaped crossover similar to the Hall coefficient. In fact, recent calculations for a Kondo lattice predict such a feature for both transport properties \cite{Coleman2005, Kim2009a}. Very recently it was shown that the coincidence of the KB and the AF QCP in \YRS\ can be removed by the application of chemical pressure which was realized by partial substitution of Rh with isoelectronic Co or Ir \cite{Friedemann2009}. In Yb systems pressure stabilizes the magnetism which manifests itself by an increase of \TN\ \cite{Friedemann2009c}. Here we report detailed resistivity measurements on \YCdS\ which lies within the range of Co-content considered in Ref. \cite{Friedemann2009}.

%
%
\section{Experimental Results and Discussion}

\begin{figure}%
	\includegraphics[width=.485\textwidth]{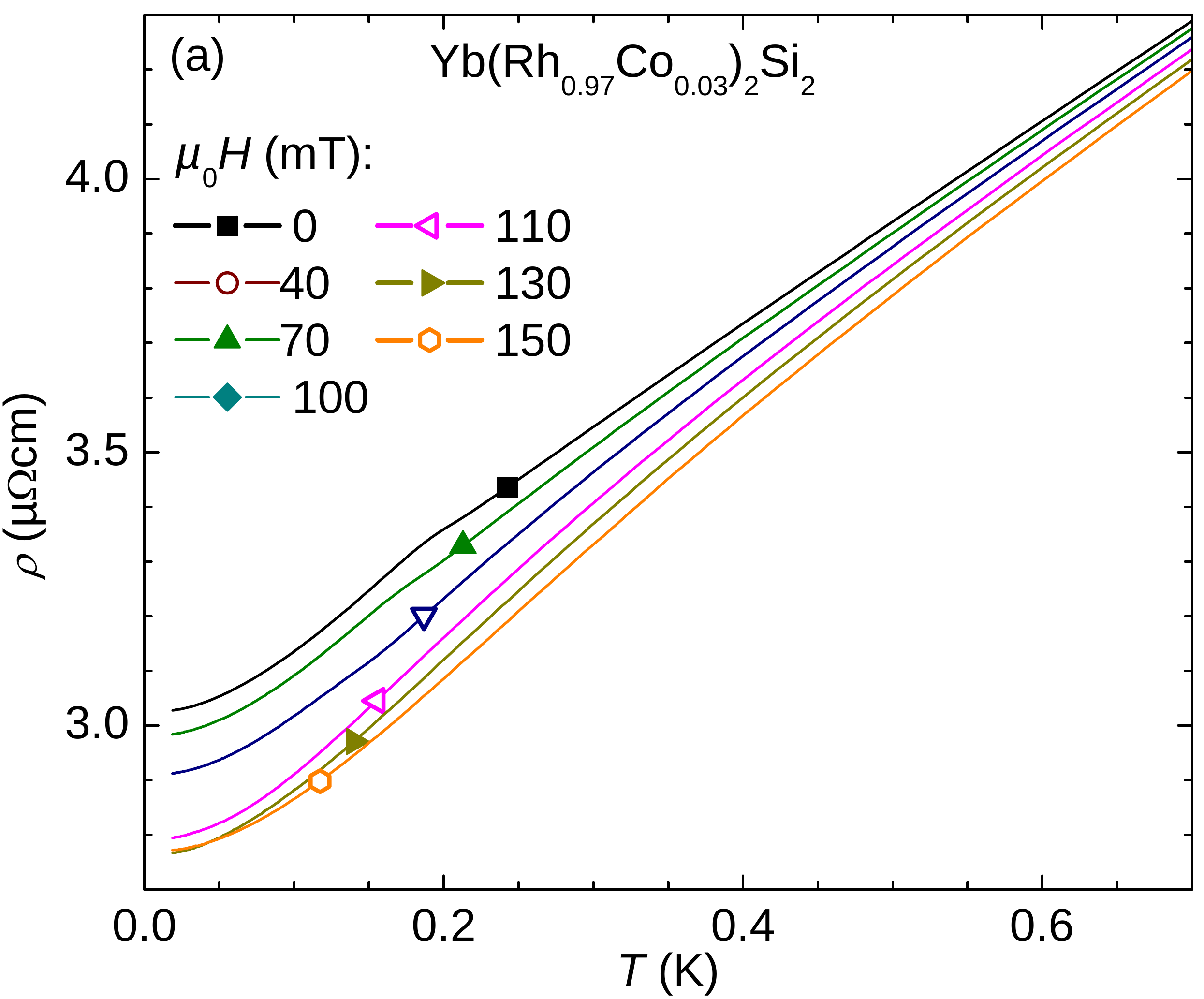}%
	\hfill
	\begin{minipage}[b]{.48\textwidth}	
		\centering
		\includegraphics[width=.68\textwidth]{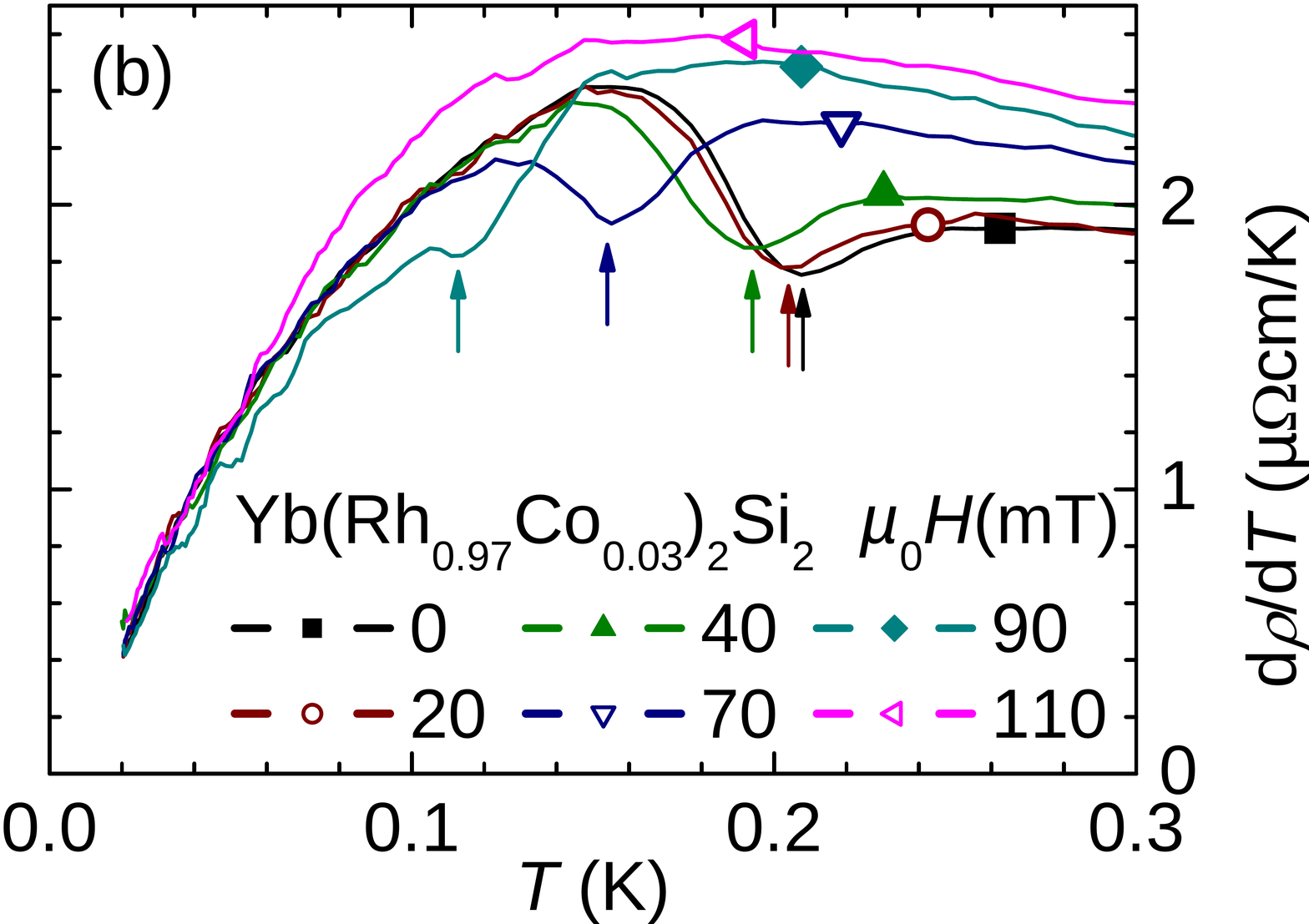}
		\caption{Temperature dependence of the resistivity $\rho(T)$ (a) and its derivative (b) for \YCdS\ at different magnetic fields. The arrows in (b) mark the local minima reflecting $\TN(H)$.}%
		\label{fig:RhovsT}%
	\end{minipage}
\end{figure}

Figure \ref{fig:RhovsT} depicts the temperature dependence of the resistivity $\rho(T)$ which is almost linear below \unit{5}\kelvin\ (cf.\ Ref.\ \cite{Friedemann2009c}). A clear drop manifests the transition to the AF ordered state with $\TN = \unit{0.21}\kelvin$ determined from the minimum in $\dd\rho/\dd T$ (Figure \ref{fig:RhovsT}b). In an external magnetic field the transition is shifted to lower temperatures and is absent for fields exceeding \unit{100}\mT.

\begin{figure}%
\includegraphics[width=.57\textwidth]{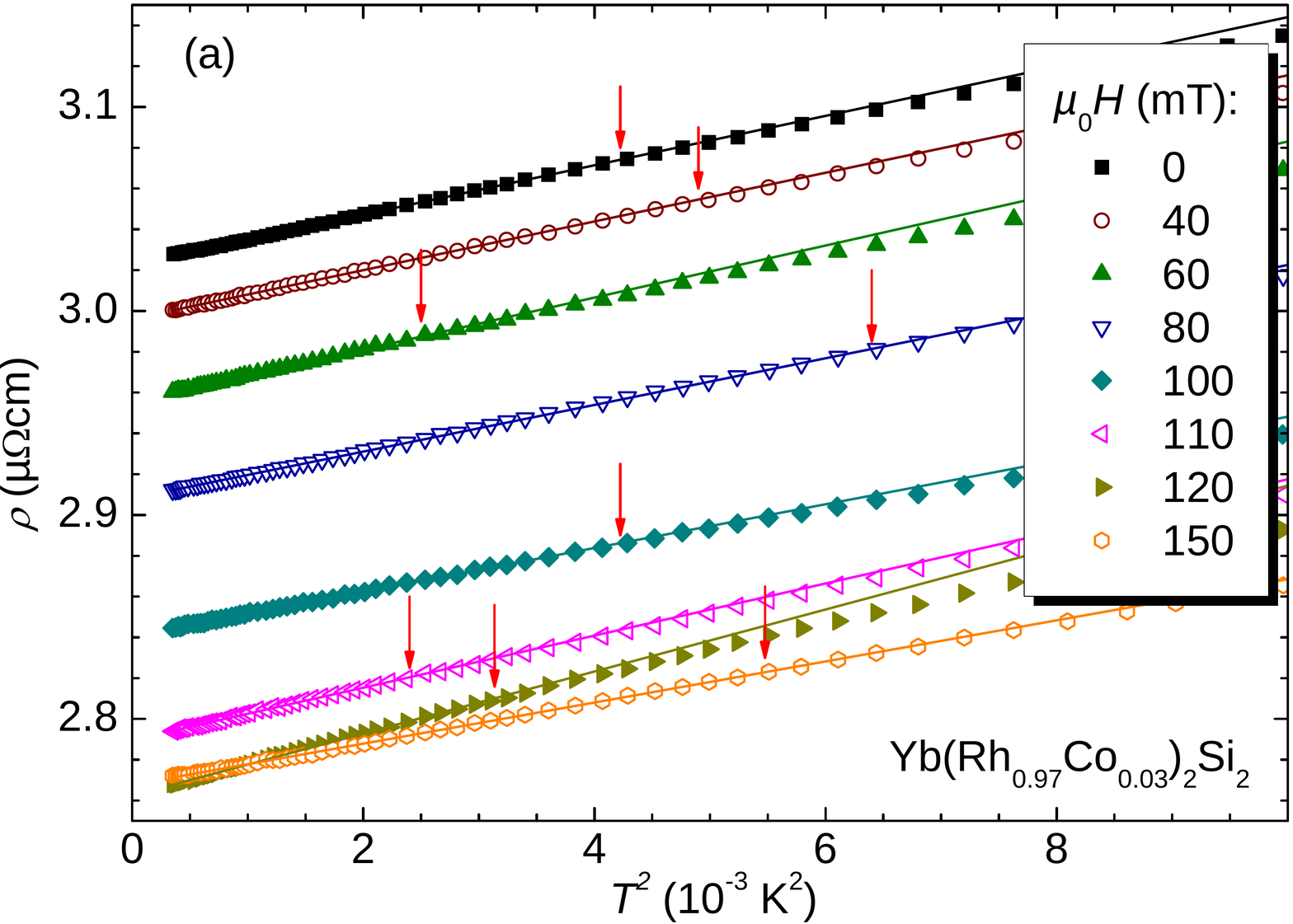}%
\hfill
\includegraphics[width=.403\textwidth]{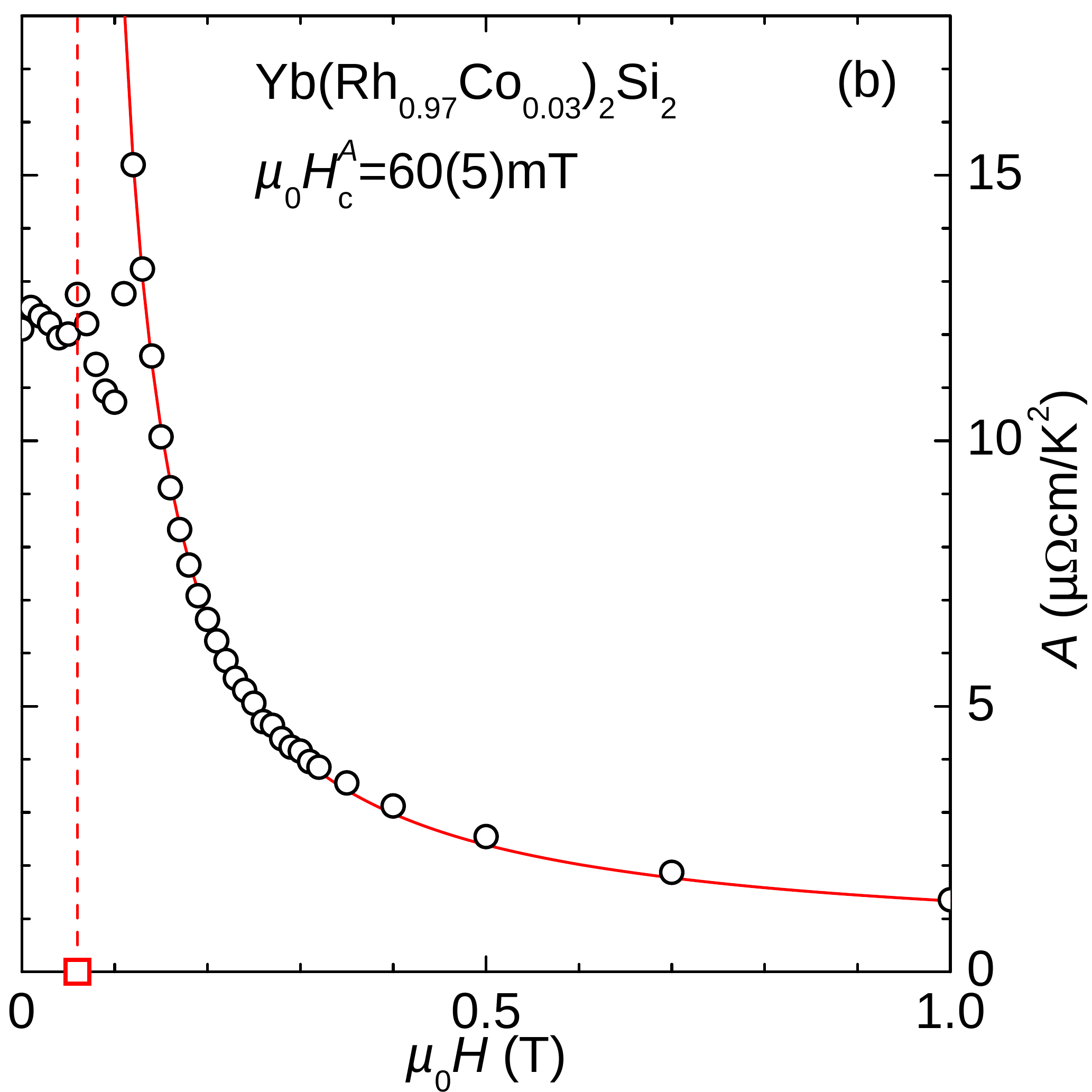}
\caption{(a) Resistivity $\rho$ plotted vs.\ $T^2$. The lines are fits of eq.\ \ref{eq:LFL} to the data at lowest temperatures. The arrows indicate the temperature \TFL\ above which the relative deviation between the fits and the data raises above \unit{0.1}\percent. The field dependence of the coefficient $A$ in (b) is depicted in the field range up to \unit{1}\tesla\ in order to highlight the behavior in the antiferromagnetically ordered phase. Here, the curve represents a fit of eq.\ \ref{eq:AKoeff} performed up to \unit{3}\tesla\ with the constant offset of $A_0 = \unit{0.4}\micro \ohm \usk \centi \metre \per \kelvin \squared$.
}%
\label{fig:RhovsTsq}%
\end{figure}

\begin{figure}%
\includegraphics[width=.57\textwidth]{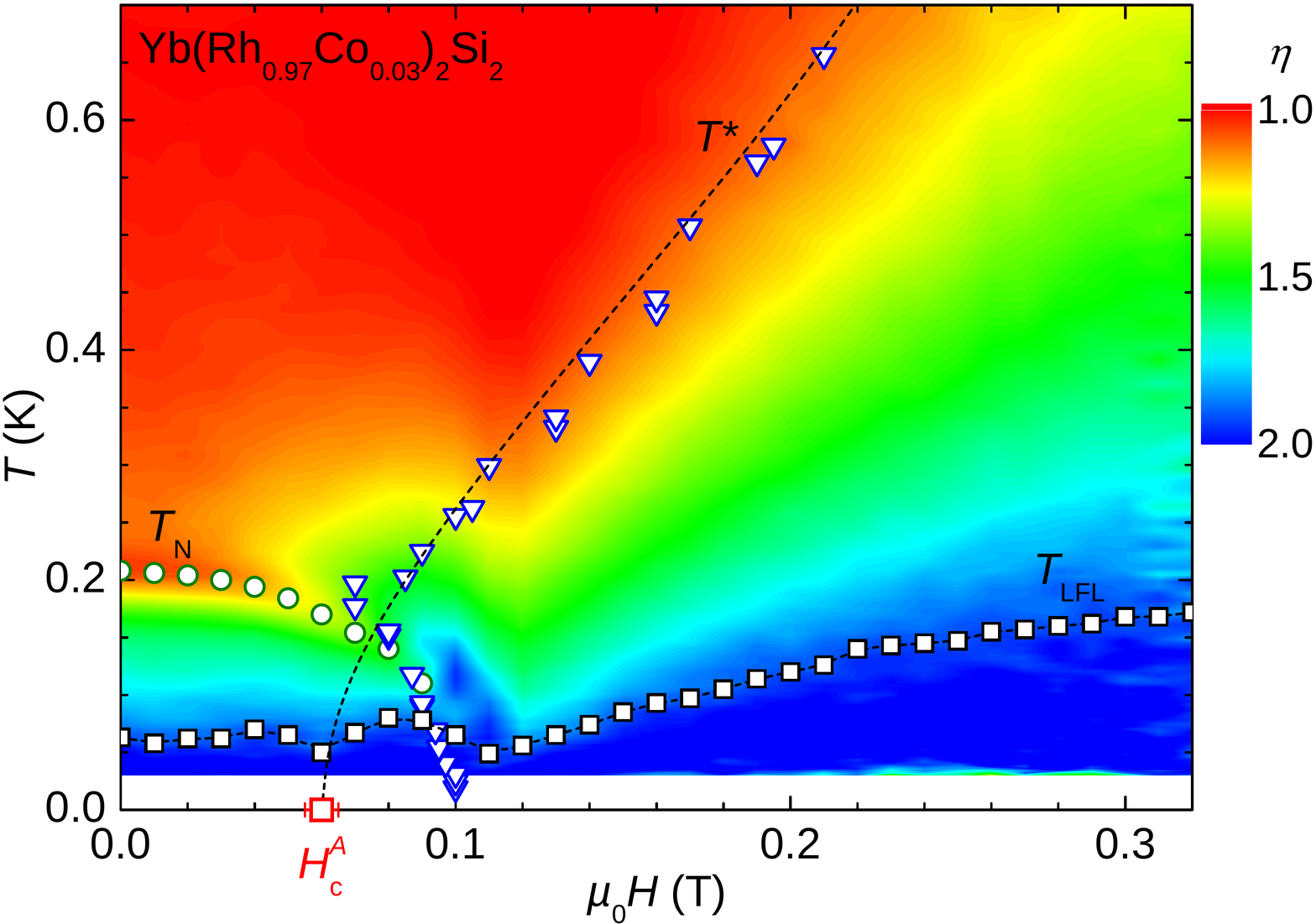}
\hfill
\begin{minipage}[b]{.43\textwidth}
	\caption{The $T$-$H$ phase diagram is constructed from the results of $\rho(T)$ and $\rho(B)$: \TN\ represents the minimum in $\dd \rho / \dd T$, $\Tstar(B)$ marks the position of the inflection in $\rho(B)$, and \TFL\ is the upper limit of the $T^2$-form in $\rho(T)$. On the abscissa \HcA\ (\includegraphics[height=.6em]{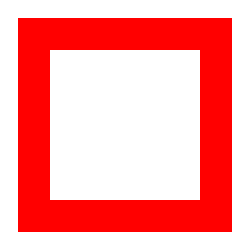}) is included. 
	The dashed line is a guide to the eye connecting \HcA\ with the inflection points of $\dd \rho / \dd T$ above \TN, thus, emphasizing the evolution of \Tstar\ as discussed in the text.
	The color code reflects the exponent of the temperature dependence of the resistivity calculated as $\eta = \dd\log(\rho-\rho_0)/\dd\log T$.}%
\label{fig:PD}%
\end{minipage}
\end{figure}

The plot of the resistivity against $T^2$ in Figure \ref{fig:RhovsTsq}a reveals the Landau-Fermi-liquid (LFL) form 
\begin{equation}
	\rho=\rho_0+A\usk T^2
\label{eq:LFL}
\end{equation} to be valid below \TFL\ (marked with arrows) at all magnetic fields applied. At small fields, \TFL\ varies between \unit{50}\mK\ and \unit{70}\mK\ with minima at \unit{60}\mT\ and \unit{110}\mT\ as can bee seen from the phase diagram in Figure \ref{fig:PD}. For fields larger than \unit{110}\mT\ $\TFL(B)$ increases monotonically.

The field dependence of $A$ (reflecting the quasiparticle-quasiparticle scattering rate) is depicted in Figure \ref{fig:RhovsTsq}b. As found for \TFL, two regimes are present: At fields below \unit{110}\mT, $A$ is almost constant with some scattering around \unit{12}\micro\ohm\usk\centi\metre\per\kelvin\squared\ whereas a strong increase is only observed when approaching from high fields which is very well described by a reciprocal divergence of the form 
\begin{equation}
	A-A_0\propto(H-\HcA)^{-1}
\label{eq:AKoeff}
\end{equation} yielding a critical field of $\HcA \approx \unit{60}\mT$. The small constant offset $A_0$ is necessary to describe the data in the entire field range up to \unit{3}\tesla\ investigated. However, the critical field is only slightly altered if the offset is omitted. 
Although the $T^2$-form is present at all fields, two regimes can be discriminated according to the behavior of $A$ and \TFL. This resembles the observations in \YRS\ where both the AF and the LFL state obey $\rho-\rho_0 \propto T^2$ \cite{CT}. However, in \YCdS\ the two regimes seem to be connected which might be due to disorder effects or due to a missing of the critical field possibly being slightly away from the applied field.

\begin{figure}%
\hfill
\includegraphics[width=.5\textwidth]{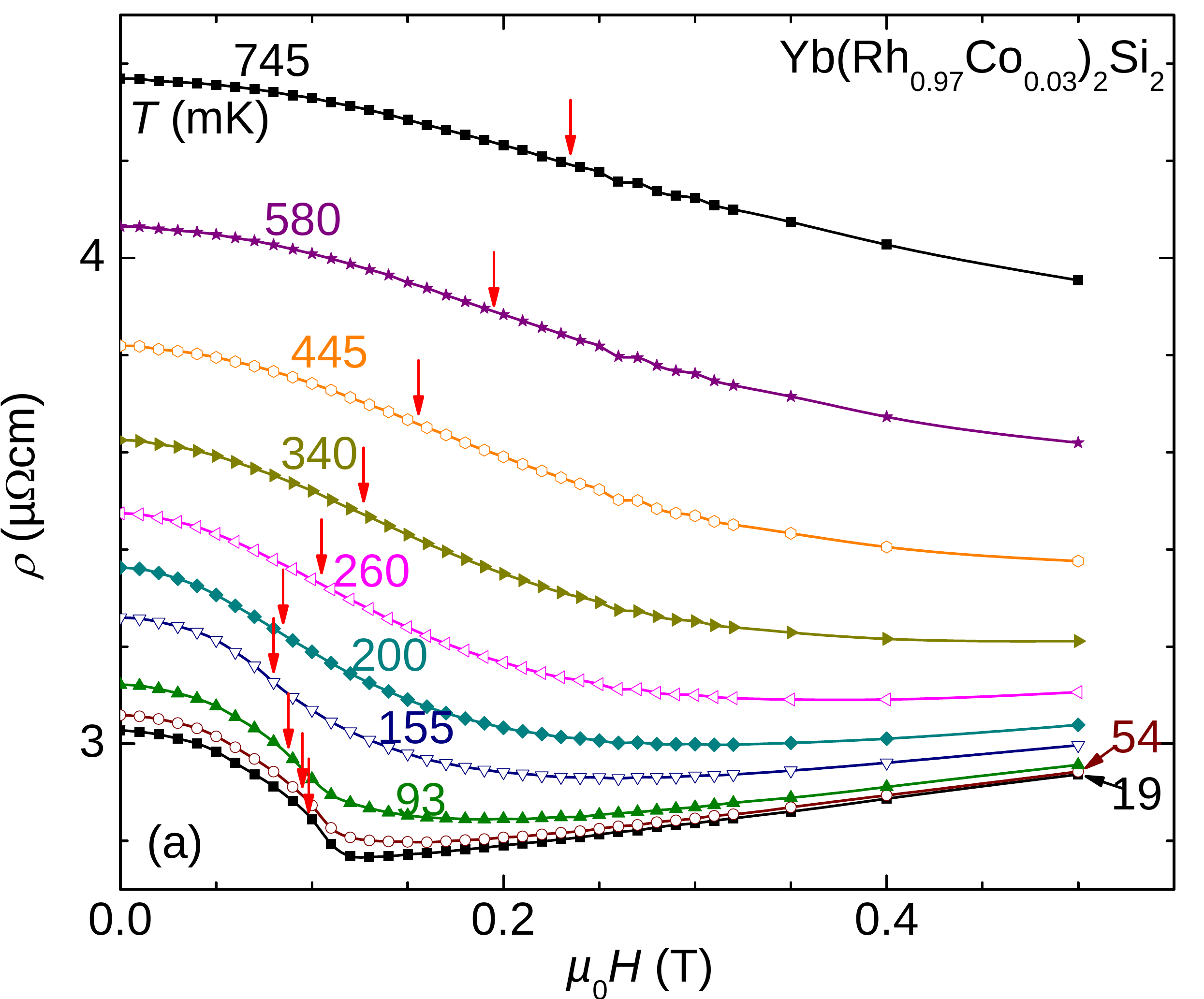}%
\hfill
\includegraphics[width=.413\textwidth]{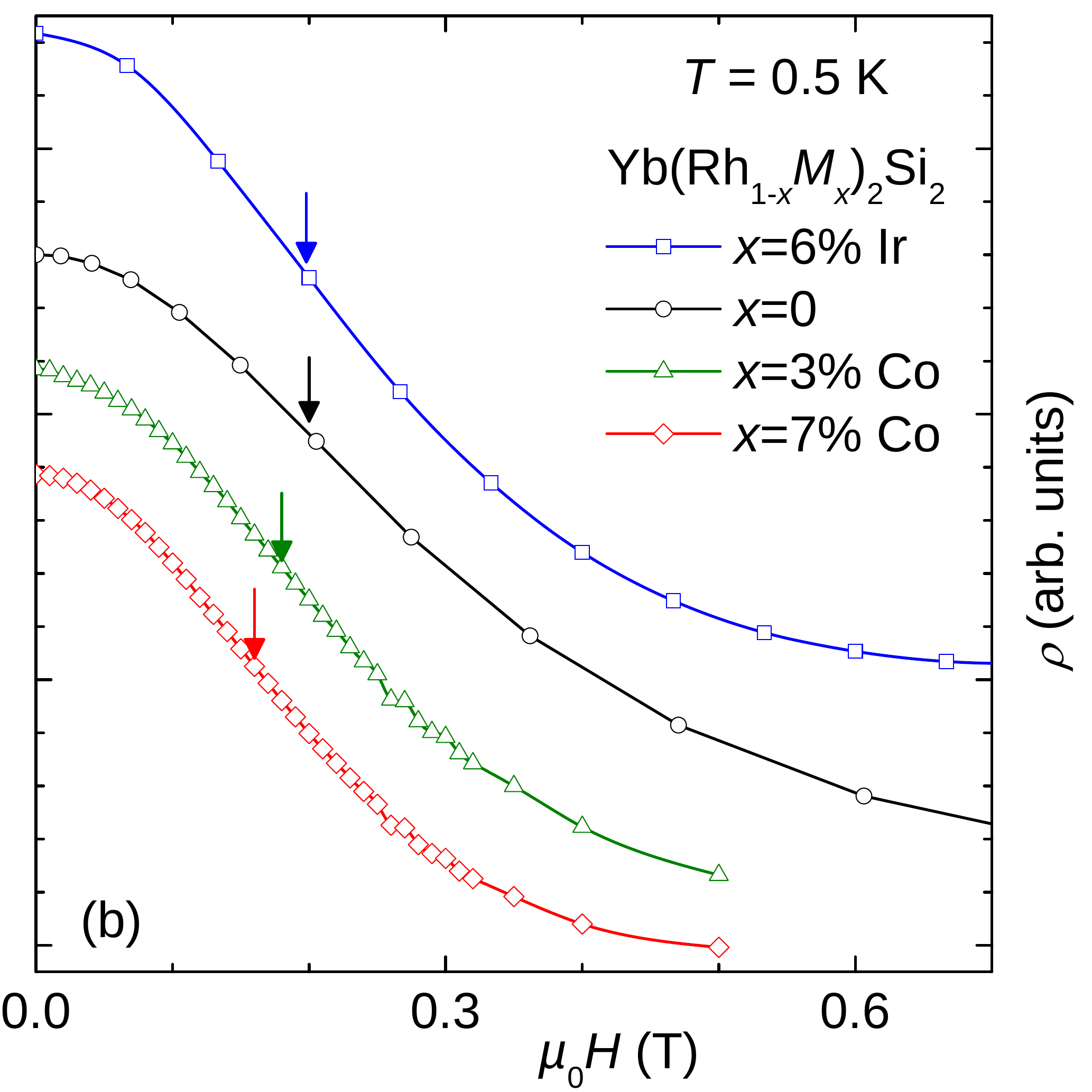}
\hfill
\caption{The magnetoresistance $\rho(H)$ of \YCdS\ at selected temperatures with the arrows indicating the inflection point. In (b) the crossover of \YCdS\ is compared to \YRS\ and other samples of the substitution series \YCxS\ and \YIyS\ at \unit{0.5}\kelvin, \textit{i.e.}\ above \TN\ where the crossover is assigned to the energy scale \Tstar.}
\label{fig:RhovsB}%
\end{figure}

Figure \ref{fig:RhovsB} shows the magnetoresistance at selected temperatures which obeys a S-shaped crossover from its zero-field value to a reduced value at elevated fields. At lowest temperatures this crossover is followed by a monotonic increase. The inflection points of the various isotherms determined from $\dd \rho / \dd H$ are included in the phase diagram in Figure \ref{fig:PD}. Below \TN, the inflection point appears to be locked to the phase boundary as it agrees with $\TN(B)$ derived from $\dd\rho/\dd T$. This allows for a precise determination of the critical field $\mu_0\HN=\unit{100(2)}\mT$. Above the \Neel-temperature, the inflection field is found to increase linearly with temperature. Obviously, in this temperature range, the magnetoresistance crossover can be assigned to the energy scale \Tstar. In fact, the direct comparison of the isothermal magnetoresistance with that of \YRS\ as well as other members of the substitution series \YCxS\ and \YIyS\ in Figure \ref{fig:RhovsB}b highlights that the position and the shape of the crossover remain very similar.

The various characteristics are collected in the phase diagram in Figure \ref{fig:PD}. Here, it becomes apparent that $\Tstar(B)$ hits the phase boundary $\TN(B)$ at finite temperatures. 
Assuming a similar curvature as observed in \YRS\ and \YIsS\ \cite{Friedemann2009}, a critical field of $\mu_0\Hstar = \unit{60}\mT$ is obtained. This matches \HcA\ and the local minimum of $\TFL(B)$. Consequently, several indications support an intersection of \TN\ and \Tstar, this way perfectly fitting into the global phase diagram of the substitution series \YCxS\ and \YIyS\ \cite{Friedemann2009}.

In addition, the exponent $\eta$ of the temperature dependence of the resistivity is shown in Figure \ref{fig:PD} as a color-coded contour. Obviously, \Tstar\ separates two regions: On the left hand side, the non-Fermi-liquid (NFL)  behavior corresponding to $\eta =1$ is present above \TN. On the right hand side $\eta$ crosses over to a value of 2 below \TFL. This indicates that the NFL-behavior is tight to the KB energy scale \Tstar\ like in \YIsS\ \cite{Friedemann2009}.
%
%
\section{Conclusion}
The results on \YCdS\ show that the magnetism is stabilized as expected. As a consequence, the AF QCP at \HN\ is shifted to higher fields compared to \YRS. By contrast, the KB energy scale \Tstar\ is not altered yielding an intersection of $\Tstar(B)$ and $\TN(B)$. A divergent $A$ coefficient indicates the presence of a QCP which is connected with \Tstar\ rather than \HN.
Finally, the critical fields \HN\ and \Hstar\ fit well into the evolution observed under chemical pressure \cite{Friedemann2009}.

\ack
Fruitful discussions with P. Gegenwart, T. Westerkamp, M. Brando and S. Wirth as well as partial support by the DFG Research Group 960 ``Quantum Phase Transitions'' are acknowledged.

\section*{References}


\providecommand{\newblock}{}

\end{document}